\documentclass[12pt]{article}

\pdfoutput=1 
\usepackage{jheppub} 

\usepackage[utf8]{inputenc}
\usepackage{jheppub}
\usepackage{epsfig}
\usepackage{bbm}
\usepackage{bbding}
\usepackage{amsmath}
\usepackage{amsfonts}
\usepackage{amssymb}
\usepackage{mathtools}
\usepackage{dsfont}
\usepackage{bm}
\usepackage{graphicx}
\usepackage{longtable}
\usepackage{pdflscape}
\usepackage{subcaption}
\usepackage{color}
\usepackage{psfrag}
\usepackage[capitalise]{cleveref}
\usepackage{hyperref}
\usepackage{tikz}
\usetikzlibrary{positioning,arrows.meta}
\usepackage{pgfplots}
\usepackage{pgfplotstable}
\usepackage{subfiles}
\usepackage{longtable}
\usepackage{tabularx}
\usepackage{ltablex}
\usepackage{booktabs}
\usepackage[export]{adjustbox}
\usepackage{listings}
\usepackage{multirow} 
\usepackage{cancel,slashed}
\usepackage{bbm}
\usepackage{graphicx}
\usepackage{latexsym}
\usepackage{color}
\usepackage{transparent}
\usepackage{mathtools}
\usepackage{array}
\usepackage{makecell}
\usepackage{graphics,psfrag}
\usepackage{placeins}
\usepackage{nowidow}
\usepackage{listings}
\usepackage[normalem]{ulem}
\usepackage{environ}
\usepackage{tikz}	
\usetikzlibrary{positioning,trees,decorations.pathmorphing,decorations.markings,decorations.pathreplacing,calc,shapes,shapes.geometric,shapes.symbols,patterns,arrows}
\pgfplotsset{
        compat=1.9,
        compat/bar nodes=1.8,
    }

\def\ls{\mathrel{\lower4pt\vbox{\lineskip=0pt\baselineskip=0pt
			\hbox{$<$}\hbox{$\sim$}}}}
\def\gs{\mathrel{\lower4pt\vbox{\lineskip=0pt\baselineskip=0pt
			\hbox{$>$}\hbox{$\sim$}}}}
\def\drawbox#1#2{\hrule height#2pt
	\hbox{\vrule width#2pt height#1pt \kern#1pt
		\vrule width#2pt}
	\hrule height#2pt}

\def\Asym#1#2{\vcenter{\vbox{\drawbox{#1}{#2}
			\kern-#2pt       
			\drawbox{#1}{#2}}}}

\newcommand{\be}{\begin{equation}}
	\newcommand{\ee}{\end{equation}}
\newcommand{\bea}{\begin{eqnarray}}
	\newcommand{\eea}{\end{eqnarray}}

\newcommand{\gsim}{\lower.7ex\hbox{$\;\stackrel{\textstyle>}{\sim}\;$}}
\newcommand{\lsim}{\lower.7ex\hbox{$\;\stackrel{\textstyle<}{\sim}\;$}}

\newcommand{\ben}{\begin{enumerate}}
	\newcommand{\een}{\end{enumerate}}
\newcommand{\bei}{\begin{itemize}}
	\newcommand{\eei}{\end{itemize}}
\newcommand{\mc}{\mathcal}

\newcommand{\coma}{\, , \quad}
\newcommand{\fstop}{\, .}
\DeclareMathOperator{\U}{U}
\definecolor{dkblue}{HTML}{BC658D}
\definecolor{lgblue}{HTML}{F5A7A7}

\begin{document}
	\pagestyle{plain}

	\makeatletter
	\@addtoreset{equation}{section}
	\makeatother
	\renewcommand{\theequation}{\thesection.\arabic{equation}}
	\pagestyle{empty}

\rightline{DESY-22-098}
\vspace{1.5cm}

\begin{center}
	\LARGE{\bf Axionic Festina Lente\vspace{2mm}}\\
	\large{Veronica Guidetti,\textsuperscript{1} Nicole Righi,\textsuperscript{1} Victoria Venken,\textsuperscript{2} \\Alexander Westphal\textsuperscript{1}\\[4mm]}
	\footnotesize{
	\textsuperscript{1}Deutsches Elektronen-Synchrotron DESY, Notkestr. 85, 22607 Hamburg, Germany}\\
	\textsuperscript{2}Institute for Theoretical Physics, Heidelberg University,\\
	Philosophenweg 19, 69120 Heidelberg, Germany\\
	\footnotesize{\href{mailto:veronica.guidetti@desy.de}{veronica.guidetti@desy.de}, \href{mailto:nicole.righi@desy.de}{nicole.righi@desy.de}, \href{mailto:victoria.venken@gmail.com}{victoria.venken@gmail.com}, \href{mailto:alexander.westphal@desy.de}{alexander.westphal@desy.de}}

\vspace*{10mm}

\small{\bf Abstract} 
\\[4mm]
\end{center}
\begin{center}
\begin{minipage}[h]{\textwidth}
The swampland conjecture known as Festina Lente (FL) imposes a lower bound on the mass of all charged particles in a quasi-de Sitter space. In this paper, we propose the aFL (axionic Festina Lente) bound, an extension of FL to axion-like particles arising from type II string theory. We find that the product of the instanton action and the axion decay constant is bounded from below by the vacuum energy. This is achieved indirectly, using dimensional reduction on Calabi-Yau threefolds, and translating the FL result for dipoles into a purely geometric bound. We discuss axionic black holes evolution, and aFL constraints on Euclidean wormholes, showing that the gravitational arguments leading to the FL bound for $\U(1)$ charged particles cannot be directly applied to axions. Moreover, we discuss phenomenological implications of the aFL bound, including constraints on string inflation models and the axion-photon coupling via kinetic mixing.

\end{minipage}
\end{center}

	\newpage
	\setcounter{page}{1}
	\pagestyle{plain}
	\renewcommand{\thefootnote}{\arabic{footnote}}
	\setcounter{footnote}{0}
	
	\tableofcontents

\section{Introduction}
Pseudo-scalar fields, called axions, were introduced into Physics beyond the Standard Model of particle physics as part of the Peccei-Quinn mechanism solving the strong CP-problem of QCD~\cite{Peccei:1977hh,Weinberg:1977ma,Wilczek:1977pj}. Since axions are derivatively coupled, they enjoy a perturbatively exact shift symmetry. Axion potentials thus arise only from non-perturbative quantum effects like instantons (breaking the shift symmetry to a discrete subgroup), or if their shift symmetry gets softly and spontaneously broken. As these axion potentials remain largely protected from uncontrolled perturbative quantum corrections, axions have become candidates for driving slow-roll inflation~\cite{Baumann:2014nda}, comprising (a part of) cold dark matter~\cite{Preskill:1982cy,Abbott:1982af,Dine:1982ah,Ringwald:2012hr,Marsh:2015xka}, and more recently for ultralight `fuzzy' dark matter~\cite{Hui:2016ltb,Marsh:2021lqg}.

The phenomenological appeal of axions motivates searching for their UV completions into quantum gravity, and most concretely, string theory as our so far best understood candidate theory of quantum gravity. UV models of axion physics are built on recent progress in constructing flux compactifications of string theory, which stabilize the moduli deformation scalar fields of the extra dimensions~\cite{Giddings:2001yu,DeWolfe:2005uu}. These models are successful in e.g. discovering mechanisms to generate large-field high-scale inflaton potentials from axion monodromy~\cite{Silverstein:2008sg,McAllister:2008hb}, or providing candidate axions able to explain sizable fractions of the observed dark matter content as fuzzy dark matter~\cite{Cicoli:2021gss}. However, the effective field theory (EFT) parameters that such string models (string vacua) need in order to provide viable axion physics often lead to backreaction. This is triggered by the other low-lying states of a given string model due to a lack of parametric scale separation.

This backreaction is not always harmful to the viability of a model, but it often appears in connection with the observation that many string vacua seem to satisfy certain constraints on their EFT properties. For instance, there appears to be a limit to the ability to send the charge of light charge carriers of gauge fields to zero for all massive charged states (called the `Weak Gravity Conjecture' -- WGC~\cite{Arkani-Hamed:2006emk}), and exact global symmetries seem to be absent~\cite{Banks:2010zn}. Such constraints, if proven to be exact consequences of e.g. string theory, would be useful in separating the space of possible low-energy EFT models into those with the chance for UV completion in string theory (the so-called `string landscape') and their complement (the `swampland')~\cite{Vafa:2005ui}.

The goal of the swampland program then is to constrain effective theories at low energy by demanding that they UV-complete into a consistent theory of quantum gravity. In this context, the electric WGC has become one of the pillars of the swampland program due to support from string theory and AdS/CFT arguments~\cite{Montero:2018fns, Harlow:2022gzl}, and its potential phenomenological reach. In $4d$ in its crudest form it states that for electrically charged black holes to be able to decay consistently, our theory must have a state with mass $m$ and charge $q$ under a $\U(1)$ gauge field with coupling $g$ such that $m < \sqrt{2} g q M_{P}$, with $M_{P}$ the $4d$ Planck scale.

In \cite{Montero:2019ekk,Montero:2021otb}, it was discovered that for charged black holes in quasi-de Sitter space with Hubble parameter $H$, to decay consistently one expects an additional constraint called the Festina Lente (FL) bound. This bound states that all charged particles must obey
\begin{equation}
    m^2 > \sqrt{6} g q M_{P} H\fstop
\end{equation}

In the years since its conception, the WGC has been extensively studied and extended in various directions (see~\cite{Harlow:2022gzl} for a recent review). In particular, it has been argued that the WGC does not just apply to gauge fields with a two-form field strength, but also to axions \cite{Arkani-Hamed:2006emk,Heidenreich:2015nta}. For the case of one axion, the WGC would constrain the axion decay constant $f$ and the action $S$ of at least one instanton charged under the axion as $S f \lesssim M_{P}$.

It is then natural to ask whether the FL bound can similarly be extended to axions. The goal of this paper is twofold. First, in Section \ref{sec:derivation}, we will argue that an extension of the FL bound to axions is consistent under dimensional reduction. We provide an argument based on a purely geometrical equivalence, starting from the original FL bound for particles. Our claim is that all instantons with action $S$ coupled to an axion with decay constant $f$ must obey
\begin{equation}
S f \gtrsim \sqrt{M_P H} \sim V^{1/4}\coma
\end{equation}
with $V$ the vacuum energy density.\footnote{It seems like the vacuum energy scale $V^{1/4}$ might serve as an IR cut-off scale in de Sitter space more broadly. For instance, \cite{Noumi:2019ohm} has suggested that the mass scale of strings might potentially be bounded from below by $V^{1/4}$.} We call this bound the axionic FL (aFL) bound. We then attempt to find a direct argument for this bound considering the consistency of axionically charged objects. However, we are unable to find such a direct, physical argument from either axionically charged black holes (\cref{sec:axionBH}) or Euclidean wormholes (\cref{sec:wormholes}). 

Second, in Section \ref{sec:phenoimplications}, we consider the phenomenological implications of the aFL bound. This is especially interesting for the early-universe cosmology, as during inflation the Hubble parameter $H$ would have been much larger than it is today.\footnote{Some implications of the original particle FL bound for inflation have been discussed in \cite{Montero:2019ekk,Montero:2021otb}. The impact of the particle FL bound on the Higgs vacuum structure and inflation was discussed in \cite{Lee:2021cor}.} We discuss how the combination of the WGC for axions and the aFL bound constrains the amplitude of periodic instanton corrections to stringy axion monodromy inflation and thus in turn the signal strength of oscillatory contributions to the CMB power spectrum and resonant non-Gaussianity. Moreover, since the WGC and aFL bounds for axions imply constraints on geometric quantities of the extra dimensions of string compactifications, we use them to constrain the parameter space of models of K\"ahler moduli (`blow-up') inflation~\cite{Conlon:2005jm}. Finally, we derive a bound for the axion-photon coupling to the Standard Model which a generic string axion might acquire from kinetic mixing. We conclude with a discussion of our results in~\cref{sec:discussion}.

\section{Derivation of the FL Bound for Axions}
\label{sec:derivation}
The general statement for the WGC for $p$-forms with gauge coupling $e_{p;d}$ in $d$ dimensions in the absence of a dilaton background reads~\cite{Heidenreich:2015nta}
	\begin{equation}\label{eq:WGCgeneral}
		\frac{p(d-p-2)}{d-2} \,T_p^2 \leq e^2_{p;d}\, q^2 M_{P;d}^{d-2} \coma
	\end{equation}
where $T_p$ is the tension of the charged $(p-1)$-brane with integer charge $q$, and $M_{P;d}$ is the Planck mass in $d$-dimensions. Such relation is degenerate for $0$-forms (axions), hence it does not directly apply. In order to get the same statement for axions, one should rely on an indirect computation, as has been done in the literature so far~\cite{Harlow:2022gzl}. In this work, following the argument proposed in~\cite{Hebecker:2015zss}, we present another derivation of the bound on $S f$ for axions by relating the quantity $S f$ to the charge-to-mass ratio of a particle to which~\eqref{eq:WGCgeneral} applies. A similar computation was carried out in~\cite{Brown:2015iha} using T-duality. As we will show, we do not need the use of T-duality,\footnote{It is crucial that we do not rely on T-duality as a Hubble constant / positive vacuum energy enters into our bound. Were we to attempt a derivation using T-duality in type II string theory we would run into the issue that type  IIA has no known de Sitter vacua that are in the controlled IIA regime. Type IIB may have de Sitter vacua, but this is an area of active discussion. Starting from there, we would have to start from a Euclidean de Sitter solution in Euclidean IIB and T-dualize one of the external Euclidean spacetime directions to dualize between an instanton and a particle. In flat space one can compactify one of the external directions to a circle of arbitrary radius and still solve the EOMs, making T-duality straightforward. However for our Euclidean de Sitter which is a four-sphere, there is only one radius of the four-sphere which solves the EOMs and we cannot just shrink either one direction or the entire four-sphere to a stringy size in order to T-dualize.} as we will express the needed relation in terms of purely geometrical quantities which are independent of the underlying theory used. 
	
Using the same logic, we are able to extend the FL bound to axions. In $4d$, the FL bound for a particle with mass $m$ and $\U(1)$ gauge coupling $e$ reads~\cite{Montero:2019ekk}\footnote{This bound follows from the demanding that the largest allowed black holes in de Sitter space charged under a $\U(1)$ gauge field evaporate back to empty de Sitter space, rather than evolving into a big crunch singularity. These black holes are known as charged Nariai black holes \cite{Nariai,Romans:1991nq} and can roughly be thought of as black holes whose horizon radius becomes as large as the cosmological horizon radius.}
\begin{equation}
\frac{e M_P^2}{m^2}\lesssim \frac{M_p}{ H} \coma
\end{equation}
where the '$\sim$' accounts for an $\mathcal{O}(1)$ constant. 

Unfortunately, this bound does not geometrize nicely. However, the gauge theory must be at weak coupling $e<1$. This then implies the weaker bound
\begin{equation}\label{eq:FLdipole}
\frac{e^2 M_P^2}{m^2}\lesssim \frac{M_p}{ H} \coma
\end{equation}
which as we shall see does geometrize in a clean way. The bound~\eqref{eq:FLdipole} can also be directly derived from ensuring black holes in de Sitter behave consistently under FL for dipoles rather than charged particles~\cite{Montero:2021otb}. Let us explain how this works in some more detail.

The bound which we will dualize to axions is the dipole version of the FL bound. While the production of electric (magnetic) dipoles clearly cannot discharge an electrically (magnetically) charged black hole, such dipoles locally screen the electric field. A dipole with moment $\mu$ in a theory with gauge coupling $e$ gains an energy $-\mu E$ ($-\mu B$) when favorably aligned in an external electric field. One then expects an instability against rapid production of dipoles for a particle of mass $m$ when $-\mu E > m$. Filling in the field strength for the maximally charged Nariai black hole, one obtains
\begin{equation}
    \mu \lesssim \frac{m}{e M_P H}\fstop
\end{equation}
The dipole moment $\mu$ is set by $e L$, with $L$ the length scale of the dipole. As we are dealing with fundamental particles, we take this to be the Compton length of the particle $L = 1 / m$. From this, then~\eqref{eq:FLdipole} follows. The same result can be derived analogously for magnetic dipoles.

One may wonder how to think of the interactions of dipoles in terms of QFT diagrammatics. This is reviewed for neutrinos in e.g.~\cite{Giunti:2008ve,Broggini:2012df}. In a diagram, the neutrino can split into a W-boson plus a lepton which form a loop and recombine back into a neutrino at the other end of the loop. The charge particles in the loop can now couple to the photon. This provides at loop level an effective vertex coupling the neutrino to the photon. One can now consider neutrino-photon interactions using this effective vertex, describing the dipole interactions of the neutrino, including the production of neutrinos by an external photon field. If one wants to think in terms of a picture without intermediate particles, the interaction coupling a Dirac spinor dipole $\nu$ to a photon takes the form $\overline{\nu} \Lambda_\mu \nu A^{\mu}$ \cite{Giunti:2008ve,Broggini:2012df}, where the $4\times 4$ spinor matrix $\Lambda^\mu$ can be decomposed into form factors. This term is of the form $\mu_M \sigma_{\mu \nu} q^\nu$ for magnetic dipoles and $\mu_E \sigma_{\mu \nu} \gamma_5 q^\nu$ for electric dipoles, with $q^\nu$ the photon momentum. One can then consider diagrams with external photon sources producing dipoles through these couplings.

\subsection{Dimensional reduction argument}
With these preliminaries out of the way, let us now turn to the dualization of our bounds. In what follows, we first derive a geometric relation for a charged particle by wrapping a D$p$-brane on a $p$-cycle. Then, we show that we can get the same geometric quantity from a D$p-1$-brane wrapping the same cycle, hence producing an axion in the non-compact space. Therefore, following~\cite{Hebecker:2015zss}, a bound for the particle carries through to the axion via the purely geometric relation.

We start with a theory in $10d$ with D$p$-branes wrapped on a $p$-cycle $\Sigma^{p}$ of the Calabi-Yau (CY) threefold $X$, namely
\begin{equation}
   \frac{1}{4 \kappa_{10}^2}\int_{M_{4}\times X} e^{\frac{3-p}{2}\phi} F_{p+2}\wedge\star^{(E)}  F_{p+2} +\mu_p \int_{Dp \text{ on } \Sigma^p}   C_{p+1}\fstop
\end{equation}
Upon compactification on the CY, this action leads to the Maxwell theory for a charged particle in 4$d$ from the reduction of the $C_{p+1}$ gauge potential, as we show in what follows. First, we introduce a symplectic basis of harmonic $p$-forms $\omega_i$ of the $p$-th cohomology of $X$. Such basis satisfies
	\begin{equation}
		\int_X \omega_i\wedge \star \omega_j = K_{ij} \coma
	\end{equation}
where $K_{ij}$ is the metric on the space of $p$-forms and is proportional to the K\"ahler metric. Hence, we can expand the ($p+2$)-form flux and the ($p+1$)-form potential in terms of the symplectic basis as $F_{p+2}= F_2^i\wedge \omega_i$ and $ C_{p+1}=A_1^i\wedge\omega_i $. The 4$d$ action is then obtained by integrating on $X$. By defining the integral charges as  
    \begin{equation}\label{eq:integralcharge}
		q^{\Sigma^{p}}_i=\int_{\Sigma^{p}} \omega_i \coma
 \end{equation}
we can write our 4$d$ theory as
\begin{equation}
 \frac{M_P^2  \,e^{\frac{3-p}{2}\phi}}{4 V_X } \int_{M_4} K_{ij} F^i_{2}\wedge\star  F^j_{2}+ \mu_p\,   \int A^i_1 q_i^{\Sigma^p}\coma
\end{equation}
$V_X$ being the CY volume in Einstein frame.
Since only a certain linear combination of gauge fields is sourced by the particle with charge $q_i^{\Sigma^p}$,
we can define the field $A_1$ and its field strength $F_2=dA_1$ as $A_1=A_1^i K^{ij}q_j^{\Sigma^p}$ and $F_2=F_2^i K^{ij}q_j^{\Sigma^p}$. The 4$d$ action then reads
\begin{equation}
 \frac{M_P^2 |q^{\Sigma^p}|^2 \,e^{\frac{3-p}{2}\phi}}{4 V_X } \int_{M_4} F_{2}\wedge\star F_{2}+ \mu_p\,  |q^{\Sigma^p}|^2 \int A_1\coma 
\end{equation}
where we introduced the notation $|q^\Sigma|^2=K^{ij}q^\Sigma_i q^\Sigma_j$. In order to extract the 4$d$ gauge coupling, we should normalize the gauge potential such that the final action reads
\begin{equation}
    S_4 \supset \frac{1}{2 e^2} \int_{M_4} F_2\wedge\star F_2 + \int_{0-brane} A_1  \fstop
\end{equation}
Therefore, the gauge coupling should be given by
\begin{equation}
    \frac{1}{e^2}= \frac{e^{\frac{3-p}{2}\phi} M_P^2}{2 \mu_p^2 V_X |q^{\Sigma^p}|^2}\fstop
\end{equation}
The particle descending from the brane wrapped on $\Sigma^p$ has mass squared given by
\begin{equation}
    m^2= \mu_p^2\, e^{\frac{p-3}{2}\phi}\, V_{\Sigma^p}^2\coma
\end{equation}
where $V_{\Sigma^p}$ is the volume of the $p$-cycle.
Finally, the ratio of the mass of the particle and the gauge coupling reads
\begin{equation}\label{eq:geomrelparticle}
    \frac{e^2 M_P^2}{m^2}= \frac{2 V_X |q^{\Sigma^p}|^2}{V_{\Sigma_p}^2}\fstop
\end{equation}
By imposing the WGC relation~\eqref{eq:WGCgeneral} as well as the FL bound~\eqref{eq:FLdipole} for a particle in 4$d$, we have that
\begin{equation}\label{eq:windowparticle}
    \frac{1}{2}\leq\frac{e^2 M_P^2}{m^2}\lesssim \frac{M_P}{ H}\fstop 
\end{equation}
Note that in order to have a particle in 4$d$ we should wrap D$p$-branes on $p$-cycles, where $p=2$, $3$, $4$ since we are working with CY manifolds. This means in turn that we are implicitly working in type IIA, where D2- and D4-branes are present, or in type IIB with D3-branes wrapped on 3-cycles.

Our goal now is to derive a geometric relation similar to the one displayed in Eq.~\eqref{eq:geomrelparticle} but for $0$-forms. Hence, we slightly change our starting setup, and we consider the very same cycle $\Sigma^p$ wrapped this time by D($p-1$)-branes, i.e. 
\begin{equation}
   \frac{1}{4 \kappa_{10}^2}\int_{M_{4}\times X} e^{\frac{4-p}{2}\phi} F_{p+1}\wedge\star  F_{p+1} +
   \mu_{p-1} \int_{D{(p-1)} \text{ on } \Sigma^{p}}C_{p}\fstop
\end{equation}
Indeed, upon compactification on the CY $X$, we get the action of an axion in $4d$. As before, we can expand the ($p+1$)-field strength and the $p$-form gauge potential in terms of the basis as $F_{p+1}= F_1^i\wedge \omega_i$ and $ C_{p}=\theta^i\wedge\omega_i$, where the $\theta^i$ are our $0$-forms. Using the definition of integral charges in~\eqref{eq:integralcharge} and compactifying on $X$, we get in $4d$
\begin{equation}
 \frac{M_P^2  \,e^{\frac{4-p}{2}\phi}}{4 V_X } \int_{M_4} K_{ij} F^i_1 \wedge\star  F^j_1 + \mu_{p-1}\,q_i^{\Sigma^p} \theta^i \fstop
\end{equation}
In order to consider again the right linear combination of fields, we further redefine the field $\theta$ and its field strength $F_1=d\theta$ as $\theta=\theta^i K^{ij}q_j^{\Sigma^p}$ and $F_1=F_1^i K^{ij}q_j^{\Sigma^p}$. The 4$d$ action then reads
\begin{equation}
 \frac{M_P^2 |q^{\Sigma^p}|^2 \,e^{\frac{4-p}{2}\phi}}{4 V_X } \int_{M_4} F_{1}\wedge\star F_{1}+ \mu_{p-1}\,|q^{\Sigma^p}|^2 \theta\fstop 
\end{equation}
After redefining the axionic field such that the final action is canonically normalized, we have that
\begin{equation}
    S_4 \supset \frac{f^2}{2} \int_{M_4} F_1\wedge\star F_1 + \theta  \coma
\end{equation}
where the kinetic term of the axion is multiplied by the decay constant $f$, which we defined to be
\begin{equation}
    f^2= \frac{e^{\frac{4-p}{2}\phi} M_P^2}{2 \mu_{p-1}^2 V_X |q^{\Sigma^p}|^2}\fstop
\end{equation}
For an axion, the mass is replaced by the instanton action $S$ coming from the wrapped D($p-1$)-brane as
\begin{equation}\label{eq:actionDp}
    S^2 = \mu_{(p-1)}^2\,e^{\frac{p-4}{2}\phi}\, V_{\Sigma^{p}}^2\fstop
\end{equation}
Finally, we arrive at the expression for $Sf$ in terms of purely geometric quantities, namely
\begin{equation}\label{eq:geomrelaxion}
        \frac{M_P}{S f}= \frac{\sqrt{2V_X} \,|q^{\Sigma^{p}}|}{ V_{\Sigma^{p}}}\fstop
\end{equation}
Note that the r.h.s. is the same geometric ratio that we found previously for a particle (cf. Eq.~\eqref{eq:geomrelparticle}). For $p=2$, $4$, this computation is valid in type IIB, while for $p=3$ we are working in type IIA.

The main point of our computation is the following: the bound~\eqref{eq:windowparticle} is actually a bound on geometrical quantities and does not contain any information on the starting $10d$ theory, namely 
\begin{equation}\label{eq:windowgeom}
    \frac{1}{2}\leq \frac{2 V_X |q^{\Sigma^p}|^2}{V_{\Sigma_p}^2} \lesssim \frac{M_P}{ H}\fstop 
\end{equation}
Therefore, as long as the cycle is the same, we are entitled to apply these bounds on the axion as well, as we managed to express the quantity $Sf$ in the same language as the particle. This finally leads to the relation
\begin{equation}\label{eq:windowaxion}
   \frac{1}{\sqrt{2}} \leq \frac{M_P}{S f} \lesssim \sqrt{\frac{M_P}{ H}}\coma 
\end{equation}
where the lower bound is the usual WGC bound for an axion coming from dimensional reduction, while the upper bound is the new FL bound for axions.

Note again that our derivation does not rely on T-duality, but only on the fact that both the relation for the particle and the one for the axion can be expressed in terms of the same quantities of the CY manifold.




\subsection{Convex shell -- WGC and aFL}
Having found the bound for a single axion from dimensional reduction, it would be interesting to extend it to a setup where multiple axions are present, as was put forward for the WGC~\cite{Rudelius:2015xta}. Therefore, we consider a theory with $N$ canonically-normalized axion fields $\phi_i$, $i=1,\dots,N$, such that their kinetic terms are given in the canonical form. Then, the potential takes the form
 \begin{equation}
     V\sim \sum_a A_a e^{-S_a} \cos\left(\sum_i \frac{\phi_i}{f_{ai}}\right)\coma
 \end{equation}
where the index $a$ runs over the number of instantons contributing to the action. The analogue of the charge-to-mass ratio vectors is~\cite{Rudelius:2015xta}
\begin{equation}\label{eq:vectoraxion}
   \textbf{z}_{a}= (\textbf{z}_{a})_i\, \textbf{u}^i= \frac{M_P}{f_{a i} S_a}\textbf{u}^i\coma
\end{equation}
where the $\textbf{u}^i$ form an orthonormal basis of the vector space.
The WGC translates into the requirement that the convex hull spanned by the vectors $\textbf{z}_{a}$ should contain the $N$-dimensional unit ball, i.e. 
\be
||\textbf{z}_{a}|| \equiv \sqrt{\mathbf{z}_{a}\cdot\mathbf{z}_{a}}>1\fstop
\ee

\begin{figure}
    \centering
    \begin{subfigure}{0.49\textwidth}
    \centering
    \begin{tikzpicture}[scale=1]
		  \node[node distance=0cm,xshift=-1.5cm,yshift=2.2cm] {\tiny{aFL}};
		  \draw[line width=1pt,draw=dkblue,fill=lgblue!30!white] (0,0) circle (2.3) ;
		  \draw[line width=1pt,draw=dkblue,fill=white] (0,0) circle (1) ;
		  \node[node distance=0cm,xshift=-0.55cm,yshift=0.4cm] {\tiny{WGC}};
		  \draw[line width=0.5pt,gray,-Triangle] (-3,0) -- node[below, pos=0.99] {\color{black}{$z_1$}}(3,0);
		  \draw[line width=0.5pt,gray,-Triangle] (0,-3) -- node[left, pos=0.99] {\color{black}{$z_2$}} (0,3);
		  \draw[line width=1pt,black,dash pattern=on 5pt off 2.5pt] (1.42,0)--(0,1.42);
		  \draw[line width=1pt,red,dash pattern=on 5pt off 2.5pt] (1.42,0) -- (1.42,1.42) -- (0,1.42);
		  \draw[line width=1.5pt,black,-Triangle] (0,0) -- (1.39,0);
		  \draw[line width=1.5pt,black,-Triangle] (0,0) -- (0,1.39);
		  \end{tikzpicture}
     \end{subfigure}\hfill
     \begin{subfigure}{0.49\textwidth}
     \centering
    \begin{tikzpicture}[scale=1]
		  \draw[line width=1pt,draw=dkblue,fill=lgblue!30!white] (0,0) circle (1.6) ;
		  \draw[line width=1pt,draw=dkblue,fill=white] (0,0) circle (1) ;
		  \draw[line width=0.5pt,gray,-Triangle] (-3,0) -- node[below, pos=0.99] {\color{black}{$z_1$}}(3,0);
		  \draw[line width=0.5pt,gray,-Triangle] (0,-3) -- node[left, pos=0.99] {\color{black}{$z_2$}} (0,3);
		  \draw[line width=1pt,black,dash pattern=on 5pt off 2.5pt] (1.42,0)--(0,1.42);
		  \draw[line width=1pt,red,dash pattern=on 5pt off 2.5pt] (1.42,0) -- (1.42,1.42) -- (0,1.42);
		  \draw[line width=1.5pt,black,-Triangle] (0,0) -- (1.39,0);
		  \draw[line width=1.5pt,black,-Triangle] (0,0) -- (0,1.39);
		  \end{tikzpicture}
     \end{subfigure}
     \caption{Convex shell: allowed axionic states should lay inside the red area, between the inner (WGC) and the outer (aFL) circle. On the left, the charge vectors satisfy both conjectures, while on the right we have a violation of the aFL bound. Indeed, charge vectors need to be big enough for all extremal dyonic black holes (inner circle) to decay, so that the black dotted line lies outside the circle. If we also impose aFL and in a situation where the outer shell is very thin, having very big charge vectors could violate the aFL bound. In the right-hand side plot, the basis charge vectors barely satisfy aFL while the $(1,1)$ charge vector violates it.}
    \label{fig:multiFL}
\end{figure}
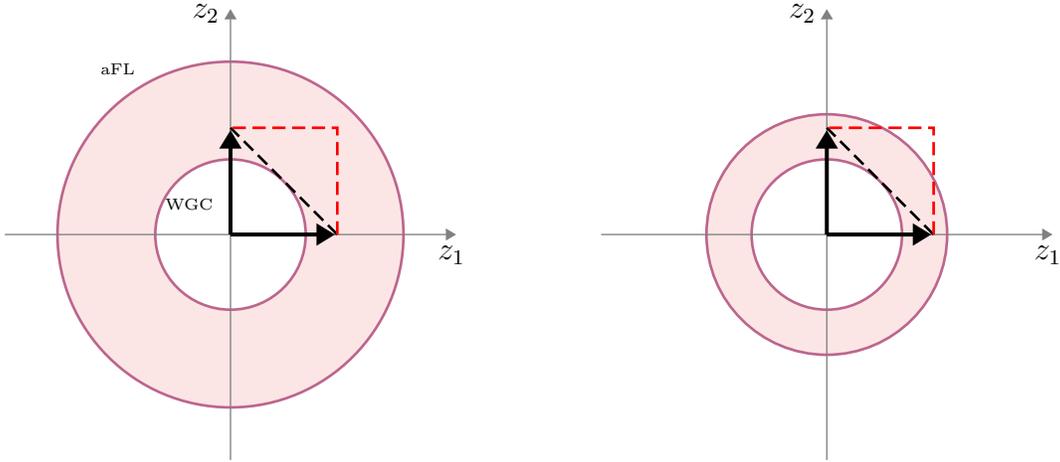

The generalization to multi-$\U(1)$s of the FL bound puts an upper bound on every vector. Consider a $\U(1)^N$ gauge theory. We denote the gauge fields with $i=1,..,N$. Let the theory have $M$ species of charged particles with masses $m_a$, $a=1,..,M$ and coupling $e_{ai}\equiv (\mathbf{e}_a)_i$ to the $i$-th $\U(1)$. The multi-$\U(1)$ version of the (dipole version of the) FL bound states~\cite{Montero:2021otb}\footnote{This is the simplest version where the $\U(1)^N$ gauge fields are not mixed. More generally, if the kinetic term for the gauge fields is $\mc{L}_{\text{kin}}=\frac{1}{4}u_{i j}F^i_{\mu \nu}F^{j\,\mu \nu}$, the bound is $m_a^2 \gg q_{ia}q_{ja}(u^{-1})^{i j}\,H M_P$ for a particle with charge $q_{ia}$ under the $i$-th $\U(1)$.}
\begin{equation}\label{eq:multiUNFL}
    m_a^2 \gg \sum_i ||\mathbf{e}_{a}||^2 \,H M_P \fstop
\end{equation}
By defining the charge-to-mass ratio vector of the $a$-th particle as $\mathbf{z}_{a}=(\mathbf{z}_{a})_i\mathbf{u}^i \equiv \left(\frac{e_{ia} M_P}{m_a}\right)\mathbf{u}^i$, we can rewrite~\eqref{eq:multiUNFL} as:
 \begin{equation}
||\textbf{z}_{a}|| <\sqrt{\frac{M_P}{H}}\fstop
\end{equation}
Hence, from~\eqref{eq:windowaxion}, we get a window of allowed values for every vector norm, namely
\begin{equation}\label{eq:windowmulti}
     1<||\textbf{z}_{a}||<\sqrt{\frac{M_P}{H}}\fstop
\end{equation}
From the derivation of the previous section, this holds also for a theory with many axions, where $\mathbf{z}_{a}$ is now given by~\eqref{eq:vectoraxion}. Note that the relation in~\eqref{eq:windowmulti} means that not only the vectors should stay outside the extremal region constrained by the WGC, but also they should lie inside the bound originating from \textit{all} the $\textbf{z}_{a}$ of the theory.

The danger then exists that if the allowed window inside the shell is very thin, the WGC convex hull will be unable to satisfy the aFL bound (see Fig.~\ref{fig:multiFL}). In particular, in presence of $N$ axions and considering the `largest' elementary axionic charge, i.e. an instanton given by a vector $\mathbf{z}_a^{\text{largest}}$ such that $\mathbf{z}_a^{\text{largest}}\cdot \mathbf{u}_i=M_P/(f_aS_a)\;\;\forall i=1\ldots N$, we have that $ ||\mathbf{z}_a^{\text{largest}}||=\sqrt{N}M_P/(f_a S_a)$. We then find that generically we must have
\begin{equation}
\label{Naxionboundaxiondep}
    \frac{N}{f_a^2S_a^2} < \frac{1}{M_P H}\coma
\end{equation}
which produces a bound on the number of allowed axions. For this bound to be very constraining, we need to have a mild hierarchy between $H$ and  $M_P$. Since $\mathbf{z}_a^{\text{largest}}$  must obey the WGC, $f_a S_a < M_P$, it then follows that
\begin{equation}
\label{Naxionboundaxionindep}
   N < \frac{M_P}{ H}\coma
\end{equation}
independently of the specific $f_a$ and $S_a$ of the largest elementary axionic charge. However, depending on the value of $f_a$ and $S_a$, \eqref{Naxionboundaxiondep} may be a significantly stronger constraint than \eqref{Naxionboundaxionindep}.

\subsection{Axionic black holes, frustrated strings and (no) big crunch}
\label{sec:axionBH}
Let us consider a black hole with non-vanishing $b=\int B_2$ charge, where the integral is over any non-trivial 2-sphere homotopic to the horizon. Such axionic black hole is described by a Schwarzschild solution~\cite{Bowick:1988xh} and the axionic charge $b$ carries no energy since the field strength $H_3=dB_2$ vanishes. Hence, if the black hole has fixed mass, it can carry arbitrary axionic charge. We want to consider now what happens if the axionic black hole starts to shrink via Hawking evaporation. Indeed, the black hole evaporates maintaining a constant axionic charge. However, if the evaporation goes through all the way down to flat space, also the sphere encircled by the Wilson line has shrunk to zero size. Clearly, this is a contradiction, and we need a way to discharge the black hole before it is completely evaporated away. 

To do so, we consider light strings `lassoing' the black hole~\cite{Hebecker:2017uix}, which couple to the field $b$ and generate an effective potential for the axion.\footnote{This can also be seen as an argument coming from requiring the absence of global symmetries in a theory of quantum gravity \cite{Banks:2010zn}.} Calling the string tension $\sigma$, as long as the radius $r$ of the black hole is larger than $1/\sqrt{\sigma}$, the strings are irrelevant and Hawking evaporation proceeds as usual. Once $r=r_b\sim 1/\sqrt{\sigma}$, the coupling of the strings to $b$ becomes sizable and in turn generates a potential for $b$ driving the axion dynamically to zero. This $b$-gradient thus forms a strong field strength $H_3$ in the near-horizon region of the black hole, which can decay by nucleating strings charged under $B_2$ in analogy to the Schwinger decay of a Maxwell electric field. In other words, when the black hole reaches a critical radius $r_b\sim 1/\sqrt{\sigma}$, the building up near-horizon field strength $H_3$ starts to decay into strings until it completely disappears. This process is expected to be almost instantaneous. The outcome of this discharging process is that we have a massive production of strings, which do not have much time to propagate and their effect is limited to a ball of radius $r_b$. However, if $r_b$ is very close to the Nariai black hole horizon size, we need to check if this string cloud provoke a big crunch singularity, as seen in~\cite{Montero:2019ekk}.

In order to understand the effects of string production, we should consider the equation of state of these light cosmic strings emitted by the axionic black hole. As we assume that they get instantaneously and massively produced, we approximate their behavior by a frustrated string network, with equation of state $p=-\rho/3$~\cite{Copeland:2009ga}. Then, we can follow the steps of~\cite{Montero:2019ekk} and study how adiabatic black hole discharging via string production would affect the fate of a Hubble patch. Let us assume that the string network can be represented as a uniform fluid and consider the FRW open slicing parametrization where the spacetime is represented as $(\mathbb{R}^2\times S^2)$:
\be
ds^2=\frac{1}{r}(dt^2- a(t)^2 dX^2)+ r^2 d\Omega^2\coma
\ee
where $r$ is the black hole radius. 
The energy conservation equation related to the frustrated string fluid is given by
\be 
\dot\rho=\rho\left(2\,\frac{\dot r}{r}+\frac{2}{3}\frac{\dot a}{a}\right) \qquad \text{implying} \qquad \rho(t)=\frac{\rho_0}{r^2 a^{2/3}(t)}\fstop
\ee
The Friedmann equations for the scale factors $a$ and $s\equiv r^2$ are
\be
\begin{array}{ll}
\displaystyle\ddot s = -\frac{1}{\sqrt{s}}\left(1-3s-\frac{\rho_0}{a^{2/3}}\right)\coma\\[10pt]
\displaystyle\frac{\ddot a}{a}= \frac{1}{2\,s^{3/2}}(1+3s)\fstop
\end{array}
\ee
Integrating the first equation and also approximating $\rho_0/a^{2/3}\sim \rho_0$ as we are mainly interested in the very beginning of the evolution (and we can fix $a(0)=1$), we find
\be 
\frac{1}{2}(\dot s)^2+V(s)=0 \qquad \text{where} \qquad V(s)=2\sqrt{s}(1-s-\rho_0)=0\fstop
\ee
We see that the maximum of $V(s)$ is located at $s=(1-\rho_0)/3<1/3$. As the size of the 2-sphere for a neutral Nariai black hole is $s = 1/3$, we see that $V(s)$ will drive the remnant of the Nariai branch away from the singularity (we are assuming that the black hole evaporates instantaneously into strings) so no big crunch is expected to happen. 

We can therefore conclude that the de Sitter black hole arguments leading to the FL conjecture for $\U(1)$ charged particles do not seem to have an obvious analog for axions. For this reason, no `pure' aFL can be constructed and the only FL constraints affecting axionic fields descend from dimensional reduction and duality, just as this is the case for the axionic WGC.

\subsection{Euclidean Wormholes}
\label{sec:wormholes}

In de Sitter space we expect the Hubble scale to give an IR cut-off for wormholes. Following along the lines of the analysis for Euclidean AdS wormholes \cite{Gutperle:2002km,VanRiet:2020pcn}, one can see that in Euclidean de Sitter space with de Sitter length $l_{dS}$ one can find a wormhole geometry
\begin{equation}
    ds^2 = \left( 1 - \frac{\tau^2}{l_{dS}^2} + \frac{c}{12}\tau^{-4} \right)^{-1} d\tau^2 + \tau^2 d \omega^2\coma
\end{equation}
where $\tau$ is the radial direction and $d \omega^2$ the metric on a three-sphere. This geometry has a smooth wormhole throat when $c<0$. We consider this regime and set $c=-|c|$. We now see that $\left( 1 - \frac{\tau^2}{l_{dS}^2} - \frac{|c|}{12}\tau^{-4} \right)$ has two physically relevant zeroes, one where the wormhole throat sits and one due to the maximal size of de Sitter. The two zeroes coincide when
\begin{equation}
\label{nariaiwormhole}
    |c| = \frac{16}{9}l_{dS}^4\fstop
\end{equation}
So when $|c| > \frac{16}{9}l_{dS}^4$ the wormhole is `too big to fit' and we obtain an upper bound on wormhole size. This is a more geometrically precise version of the statement that the wormhole neck, the radius $r_0 \sim c^{1/4}$ where the wormhole is at its narrowest, should fit inside a Hubble patch, 
\begin{equation}
\label{eq:hubblemaxthroat}
r_0 < 1/H\fstop 
\end{equation}
The preceding provides a purely geometrical IR cut-off for wormholes in de Sitter space. For Giddings-Strominger axion wormholes \cite{Giddings:1987cg}, $r_0$ depends on $f$. Therefore, aFL will provide an alternative IR cut-off for axion wormholes. We will derive this cut-off using the flat-space expressions for the Euclidean wormhole geometries as an approximation. We will see a posteriori that this approximation is justified.

With the notation of~\cite{Hebecker:2018ofv}, control on semiclassical gravity requires the neck radius to be large enough, i.e.
\begin{equation}
\label{eq:wormholefneck}
    r_0= \sqrt{\frac{1}{2 \sqrt{6}\pi^2 f M_P}}>\frac{1}{M_P}\fstop
\end{equation}
We now cast \cref{eq:windowaxion} as a constraint on $f$ as
\begin{equation}
    \frac{S}{\sqrt{2}M_P} \leq \frac{1}{f} \leq \frac{S}{\sqrt{M_P H}}\coma
\end{equation}
and substitute $r_0$ for $f$ via \cref{eq:wormholefneck} to obtain
\begin{equation}
\label{eq:wormholesizebound}
 \frac{\sqrt{S}}{2 (3^{1/4}) \pi M_P}   \leq r_0 \leq \frac{\sqrt{S}}{2^{3/4}3^{1/4}\pi M_P^{3/4}H^{1/4}}\fstop
\end{equation}
We see that, barring a very large $S$, this provides an IR-cutoff for $r_0$ at a far shorter length scale than \cref{eq:hubblemaxthroat}. Then, aFL implies that the largest axionic wormholes allowed in de Sitter space are sufficiently small that they can effectively be treated as flat space wormholes, justifying our approximation as promised.

We saw that there exists a largest geometrically allowed wormhole in de Sitter, with $|c|$ given by \cref{nariaiwormhole}. One can think of this as the wormhole analogue of the Nariai black hole. As the particle version of FL is derived from considering the decay of charged Nariai black holes, it is tempting to hope that one could derive aFL using this `Nariai wormhole'. However, one is faced with several issues which we believe make this approach not viable. 

First, since aFL implies via \cref{eq:wormholesizebound} an IR cut-off for wormholes at shorter length scales than the `Nariai wormhole', aFL implies that `Nariai wormholes' cannot exist, and it is unclear how considering the decay of such a wormhole could lead to aFL. 

Second, conceptually, the wormhole is a Euclidean instantonic object. Given the lack of a time direction it is unclear what we would mean by the wormhole decaying so rapidly that there would be a crunch-like singularity.

Third, the objects that would discharge an axionically charged wormhole would be strings, analogously to how charged particles discharged the charged Nariai black hole through the Schwinger process. One may argue that for sufficiently light strings compared to the $H_3$-flux of the wormhole, their production is unsuppressed. However, as in \cref{sec:axionBH} the fundamental issue remains that the equation of state of frustrated string networks does not drive one towards a big crunch.

One way in which one might be able to argue for aFL from wormholes is if one can provide an independent argument why large wormholes, in the sense that $r_0 >\sqrt{S}/2^{3/4}3^{1/4}\pi M_P^{3/4}H^{1/4}$, are not allowed to exist tout court. However, we are not aware of an independent argument for this.

\section{Phenomenological Implications}
\label{sec:phenoimplications}
We can apply our findings to make phenomenological predictions on inflationary models concerning axions as inflatons. Note that the very high quasi-cosmological constant during high-scale inflation imposes an electromagnetic FL bound on the SM charged matter states which all of them violate drastically. One possible way around this consists of assuming a so-called portal coupling of the SM Higgs $h$ to the SM-neutral inflaton $\phi$~\cite{Lebedev:2021xey}. This coupling $g_{h\phi}\phi^2 h^2$ with $g_{h\phi}<0$ induces a very large EW symmetry breaking Higgs VEV during inflation, thus providing an avenue to render the whole SM matter spectrum massive enough to satisfy the FL bound. There will be bounds on the size of $|g_{h\phi}|<1$ such that this portal coupling does not affect the inflationary dynamics too strongly via radiative corrections. We leave details and an analysis of the potentially interesting phenomenology of this solution for the future.

In what follows, we focus on axion monodromy inflation~\cite{Silverstein:2008sg,McAllister:2008hb} and blow-up inflation~\cite{Conlon:2005jm}, and we derive implications coming from imposing the window required by the FL bound and the WGC. Finally, we also show how the aFL bound implies a lower bound on the kinetic mixing parameter for the axion-photon coupling.

\subsection{Axion monodromy inflation}
Monodromy in the axion potential is generated when non-perturbative effects such as branes introduce non-periodic terms on top of the harmonic ones. The periodic part is dominant for small vacuum expectation values of the inflation, while it gets exponentially suppressed at large volume, and the monodromy term takes over on large field displacements. We can consider the following inflationary potential
\be
\label{eq:monodromy}
V(\phi)=\mu^{4-p}\phi^p+\Lambda^4\cos\left(\frac{\phi}{f}\right)\qquad \text{where}\qquad p\leq2\coma
\ee
where the first terms represents the monodromy-induced potential with $p\leq2$ while the second term comes from instanton corrections at a scale $\Lambda^4$. The resonant non-Gaussianities arising from this model are \cite{Flauger:2010ja}:
\be
f_{NL}^{\text{res}}=\frac{3\sqrt{2\pi}}{8} b^* \omega^{*3/2}\coma
\ee
 where $^*$ denotes evaluation at horizon exit, $\omega^*=\sqrt{2\epsilon^*} M_P/f$ is the resonance frequency, $b^*$ is a parameter measuring the instanton generated wiggles intensity in the inflationary potential,
\be
b^*=\frac{\Lambda^4}{p\, f \mu^{4-p} (\phi^*)^{p-1}}\coma
\ee
and $\epsilon$ is the slow-roll parameter derived from the monodromy potential only. 
Indeed, to match the experimental constraints on the power spectrum of density perturbation, the oscillating part of the potential must be highly suppressed at the pivot scale. The monotonicity of the potential and the consistency with observational constraints require $b^*\ll1$ for $p\geq 1$ but this condition is believed to hold in the more general case we consider. To get the range of admissible values for $f_{NL}^{\text{res}}$ for each value of $p$, we remove a model degree of freedom imposing COBE normalization (i.e. fixing the size of the curvature perturbations) as
\be
\sqrt{P_s(N^*)}\simeq\frac{1}{10\pi}\sqrt{\frac{4}{3}\frac{V_{\Lambda=0}(\phi^*)}{\epsilon^* M_P^4}}\simeq 2 \times 10^{-5}\fstop
\ee
\begin{figure}[t]
    \begin{center}
    \includegraphics[height=0.30\textheight]{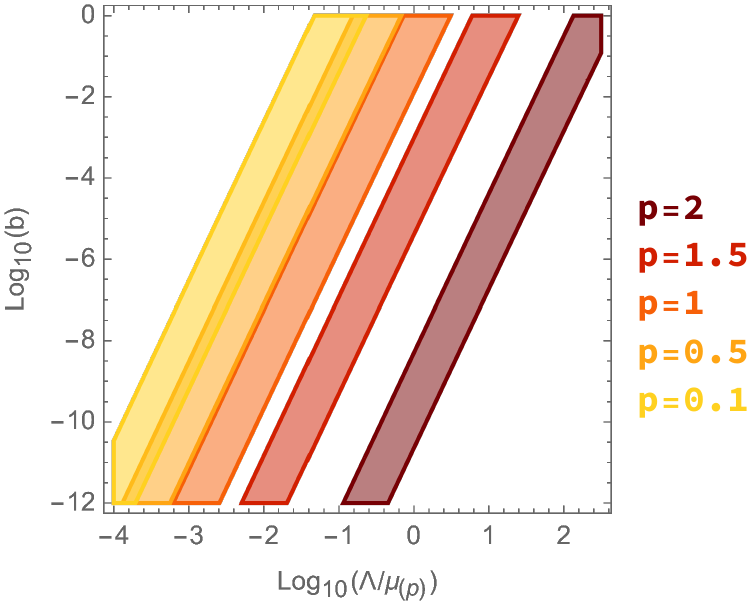}\hspace{30pt}\includegraphics[height=0.3\textheight]{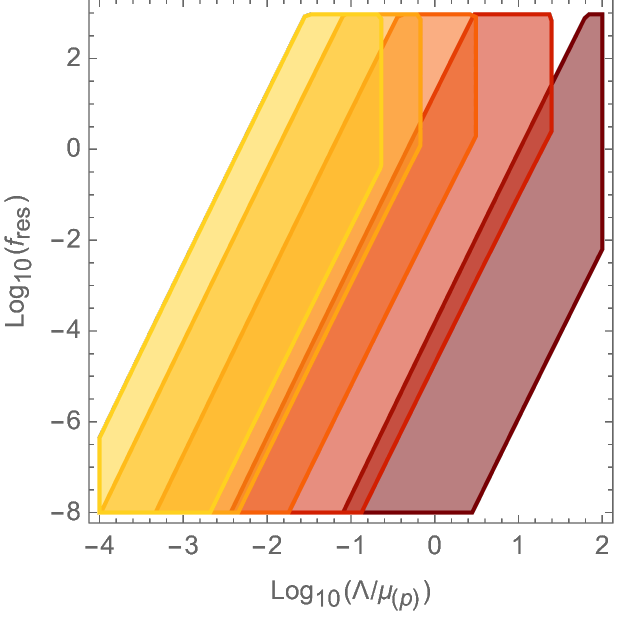}
    \caption{Theoretically allowed values of $b^*$ (left) and $f_{NL}^{\text{res}}$ (right) coming from imposing monotonicity of the inflationary potential, i.e., $b^{*}<1$, FL and WG conjectures, and requiring that $\omega^*<10^2$ to match experimental constraints \cite{Planck:2019kim}. These results refer to the monomial monodromy potential in Eq. (\ref{eq:monodromy}), fixing the number of e-folding in absence of wiggles to be $N=50$ and imposing COBE normalization.}
    \label{fig:f&bBounds}
    \end{center}
\end{figure}
Moreover, the wiggles in the potential cannot significantly affect the number of e-folding $N$ unless a massive parameter fine-tuning is performed; therefore, as experimental constraints require $40<N<60$, we fix $N=50$ when $\Lambda=0$, thus removing another degree of freedom from the model. This way, we uniquely identify $\mu$ and the value of $\phi^*$. We stress that our results do not significantly change when choosing other $N$ values within the allowed range. Using the upper and lower bounds in \eqref{eq:windowaxion}, also imposing the relation $\Lambda^4=e^{-S}$, we get to the theoretically allowed window for $b^*$ and $f_{NL}^{\text{res}}$, WGC and aFL providing the lower and upper bound respectively. We plot these results in Fig. \ref{fig:f&bBounds}.
Finally, requiring that the lower bound on $b^*$ coming from WGC is not in contrast with the condition $b^*\ll1$ sets a lower bound on the instanton action $S\gtrsim 25$. It can be easily checked that this relation is almost insensitive to the value of $p$.

\subsection{Constraints on string compactifications and blow-up inflation}
Now, let us focus on the geometric relation coming from dimensional reduction~\eqref{eq:windowgeom}. Then, we see that only the following volume relations among internal cycles are allowed:
\begin{equation}
\label{eq:geometricimplications}
    \sqrt{\frac{H}{M_P}}\lesssim\frac{V_{\Sigma}}{\sqrt{ V_X} \,|q^{\Sigma}|}=\frac{V_{\Sigma}}{\sqrt{q_i^{\Sigma} \mc{K}^{ij}q_j^{\Sigma}}}\lesssim 2\coma
\end{equation}
where $\mc{K}^{ij}=V_X K^{ij}$ is the inverse K\"aheler metric. It is easy to see that FL is pointing towards a preference for low scales, as in high scale inflation models the window of allowed values for $S f$ gets narrow. This window should be valid during all the stages of the EFT, so we could work at inflationary values for $H$, where the window is at its narrowest point. Computing these constraints in real string-inflation models leads to lower bounds for the volumes of local cycles, stating that the volume of the cycle supporting the instanton should be finite. On the other hand, this window has no impact on large cycles parametrizing the overall extra-dimension volumes, as in this case the aFL becomes a constant that naturally satisfies the l.h.s. of Eq. (\ref{eq:geometricimplications}) in realistic inflationary models. 

Let us consider, for instance, the case of K\"ahler moduli (also called blow-up) inflation \cite{Conlon:2005jm} in the context of Large Volume Scenario (LVS)~\cite{Balasubramanian:2005zx} stabilization. This model requires that the overall volume $V_X$ of the internal geometry gets stabilized via one or more small 4-cycles with volume $\sim \log V_X$. The inflaton field, $\tau_{\text{inf}}$, is played by an unstabilized blow-up modulus that appears displaced from the minimum of its potential (but whose displacement does not massively affect volume stabilization). Assuming the standard Swiss-cheese geometry, the bound on the inflaton axion partner coming from Eq. \eqref{eq:geometricimplications} reads (assuming that the instanton wraps the inflaton cycle and setting the charge equal to 1)
\begin{equation}
    \frac{H}{M_P}\lesssim\frac{\tau_{\text{inf}}^{3/2}}{ V_X}\lesssim 4\fstop
\end{equation}
We now look at the consequences of this relation on the inflationary parameters. In these models, the slow-roll parameter $\epsilon$ scales like $\epsilon\sim V_X^3 e^{-2S_{\text{inf}}}$, where $S_{\text{inf}}=a \tau_{\text{inf}}$ is the action of the instanton wrapping the inflationary cycle. Satisfying COBE normalization requires $e^{ S^*_{\text{inf}}}/V_X^{5/2}>5\cdot 10^{-4}$ ($^*$ denotes horizon exit time) so that the bound on $S^*_{\text{inf}}$ becomes
\be
S^*_{\text{inf}}>2.5 \left(\ln(V_X) - 3 \right)\fstop
\ee
Hence, the upper bound on $\epsilon$ at horizon exit reads
\be
\epsilon^*<\frac{3 \cdot 10^6}{V_X^2} \coma
\ee
which induces a bound on the tensor-to-scalar ratio, being $r=16\epsilon$. The typical overall volumes considered in K\"ahler moduli inflation are $V_X\sim 10^6$ implying $\epsilon^* < 3\cdot 10^{-6}$ and $r<5\cdot 10^{-5}$.

\subsection{Axion-photon coupling from kinetic mixing}
In gauge theories containing more than one $\U(1)$, it is possible to have kinetic mixing terms between different $\U(1)$ gauge fields \cite{Holdom:1985ag,Foot:1991kb,Dienes:1996zr}. The consequence of kinetic mixing is that matter fields charged under some $\U(1)$ also acquire a charge under another $\U(1)$ which is proportional to the kinetic mixing parameter. In the context of type II models, kinetic mixing gets naturally generated in the effective Lagrangian by one-loop effects \cite{Abel:2008ai} when massive modes coupled to different $\U(1)$s are integrated out. In order to understand the aFL conjecture induced bounds on the kinetic mixing parameter, we start by summarizing how to derive gauge kinetic functions and their dependence on moduli fields. Applications of the original FL bound on systems containing $\U(1)$ kinetic mixing between visible and hidden sectors can be found in \cite{Ban:2022jgm}.

Let us consider the DBI action (with the $B$-field set to zero) for a D7-brane wrapping a 4-cycle $\Sigma^4$. Expanding the action in powers of the gauge field strength, we get
\begin{equation}
    S_{\text{DBI}}
    = -\int_{M_4\times \Sigma^4}  \frac{\text{d}^8\xi\, e^{-\phi} \sqrt{-g}}{(2\pi)^{7}(\alpha')^{4}} \left(1+\frac{(2\pi \alpha')^2}{4} G_{\mu\nu}G^{\mu\nu}\right)\fstop
\end{equation}
By using the definition of the K\"ahler modulus in Einstein frame, i.e.
\begin{equation}
    \text{Re}(T)=\frac{e^{-\phi}}{(2\pi)^4(\alpha')^2}\int_{\Sigma^4} \sqrt{g_{\Sigma^4}}\equiv \tau \coma
\end{equation}
we get, upon dimensional reduction on $\Sigma^4$,
\begin{equation}
   S=
   -\frac{\tau}{4(2\pi)}\int_{M_4} G\wedge\star G\fstop
\end{equation}
Hence, we get the $4d$ effective action
\begin{equation}\label{eq:finalnomalizationhidden}
   S= -\frac{1}{2 g_{\text{h}}^2} \int_{M_4} G\wedge\star G \quad \text{ for }\quad g_{\text{h}}^2= \frac{4\pi}{\tau}\coma
\end{equation}
where `h' stands for hidden, as we assume this theory lives in the hidden sector.

Consider an effective Lagrangian describing two $\U(1)$ gauge fields. The electromagnetic field strength $F$ is located in the visible (`v') sector, while the other field strength $G$ lives in the hidden (`h') sector. The visible sector axion plays the role of the QCD axion while the hidden sector axion $a$ is a closed string axion. Omitting the part for the QCD axion, such a Lagrangian reads~\cite{Abel:2008ai,Goodsell:2009xc,Cicoli:2011yh}
\begin{equation}\label{eq:Lag1}
  \mathcal{L}\supset -\frac{1}{2 g_{\text{v}}^2} F'\wedge\star F' -\frac{1}{2 g_{\text{h}}^2} G'\wedge \star G' + \frac{\chi}{g_{\text{v}}g_{\text{h}}} F'\wedge\star  G'
  -\frac{a'}{8\pi^2}\, G' \wedge G'\fstop
\end{equation}
We should manipulate this Lagrangian in order to recover the canonically normalized form. First, we redefine $F=(\sqrt{2}/g_{\text{v}} )F'$, $G=(\sqrt{2}/g_{\text{h}} )G'$ and by requiring the canonical form for the axionic kinetic term 
we have $a=a' f$. By further substituting $g_{\text{h}}^2= \frac{4\pi}{\tau}$ (cf. \eqref{eq:finalnomalizationhidden}), and using the relation $\tau= SN/2\pi$ coming from $N$ branes wrapping the cycle with volume $\tau$ generating the instanton acion $S$, Eq.~\eqref{eq:Lag1} becomes
\begin{equation}\label{eq:Lag4}
  \mathcal{L}\supset -\frac{1}{4} F\wedge\star F -\frac{1}{4} G\wedge \star G + \frac{\chi}{2} F\wedge\star  G
  -\frac{a}{S f N}\, G \wedge G\coma
\end{equation}
where $N$ is the number of branes wrapping the cycle.
Now, we have to diagonalize the gauge kinetic terms via $G=\hat{G}+ \chi F$, finally obtaining 
\begin{equation}
\begin{split}
    \mathcal{L}\supset&  -\frac{1}{4}\left(1-\chi^2\right) F\wedge\star F -\frac{1}{4} \hat{G}\wedge\star \hat{G} \\&\, - \frac{a}{N Sf}\left(\hat{G}\wedge\hat{G}+\chi \hat{G}\wedge F+\chi F\wedge\hat{G}+\chi^2 F\wedge F\right)
    \fstop
\end{split}
\end{equation}
In particular, we see that $a$ acquires a coupling to the ordinary photon of the form
\begin{equation}
    \mathcal{L} \supset -g \,a\, F \wedge F\coma
\end{equation}
where we have defined the coupling $g \equiv\chi^2/(N S f)$. Note that $[g]= M^{-1}$, and in the experimental literature is given in $\text{GeV}^{-1}$. Now, we can apply the aFL bound to $g$, i.e.
\begin{equation}
   \frac{1}{S f}\lesssim \frac{1}{\sqrt{M_P H}} \quad \Rightarrow \quad g\lesssim \frac{1}{ N}\frac{\chi^2}{\sqrt{M_P H}} \coma
\end{equation}
which can also be read as 
\begin{equation}\label{eq:mixangle}
    \chi^2 \gtrsim N g\sqrt{H M_P}\fstop
\end{equation}
This result strongly depends on the starting setup, mainly on the geometry of the internal manifold. Also, we disregarded the contribution of the QCD axion and applied the single-axion version of the aFL bound for ease of exposure. A specific implementation would require detailed information about $S$ and $f$ of the instantons charged under both $\U(1)$s and make this constraint stronger. 

The only parameter in \cref{eq:mixangle} which is specific to our brane set-up is the number of branes $N$, which since it is integer obeys $N\geq 1$. Much as we were only able to derive aFL from particle FL for branes wrapping cycles in a CY three-fold but then conjectured that aFL should hold generically in any consistent theory of quantum gravity, we will now conjecture that in any consistent theory of quantum gravity the mixing angle between the axion and the photon should obey
\be
  \chi^2 > \mathcal{O}(1) g\sqrt{H M_P} \sim g V^{1/4}\,,
\ee
with some unknown $\mathcal{O}(1)$ factor. Given a specific model coming from string compactification one may attempt a derivation analogous to the derivation of \cref{eq:mixangle} to obtain a specific bound whose coefficient depends on the geometric data and may be much larger than $\mathcal{O}(1)$.

Hence, from \cref{eq:mixangle} we are able to relate the value of the coupling with the mixing angle.
The quantity $g$ is constrained by experiments to be $g<10^{-10}\,\text{GeV}^{-1}$~\cite{Payez:2014xsa} (see also~\cite{Dror:2020zru}). In order to estimate our bound, we use $g\sim 10^{-11} \,\text{GeV}^{-1}$ together with the Hubble constant valued nowadays $H_0\sim 10^{-33}$ eV and $N=1$. By plugging these values in~\eqref{eq:mixangle}, we derive a lower bound on the mixing angle $\chi \gtrsim 10^{-12} $. We point out that recently~\cite{Minami:2019ruj,Eskilt:2022wav} proposed a value for the mixing angle from the cosmic birefringence detected by \textit{Planck}~\cite{Planck:2016soo} which sits inside our bound.
    
Seen from another perspective, this bound is in tension with some proposals of kinetic mixing from model building~\cite{ParticleDataGroup:2020ssz}. Note that smaller values for $g$ relax this tension, while $N>1$ increases the lower bound for $\chi$. To estimate the consequences of having smaller $\chi$, we can use a lower value from the most recent predictions, i.e. $\chi< 10^{-16}$, which in turn from the aFL bound implies $g\lesssim 10^{-20}\,\text{GeV}^{-1}$ (keeping $N=1$).

\section{Discussion}
\label{sec:discussion}

In this work, we provide an extension of the FL bound to fundamental axions, i.e. those axions obtained via dimensional reduction of $p$-form gauge potentials of $10d$ type II string theories. The original FL bound itself states a bound on the charge-to-mass ratio of particles charged under an electromagnetic-type $\U(1)$ gauge theory in de Sitter space-time. This bound originates from the fact that Schwinger pair production in the electric field of an extremal charged Schwarzschild-de Sitter black hole of maximal size, an extremal charged Nariai black hole, would induce a space-time crunch if the FL bound on the charge-to-mass ratio were violated.

Our argument establishing an axionic Festina Lente (aFL) bound relies first on expressing the (dipole version of the) FL bound as a bound on the ratio of the internal volume and the volume of the cycle supporting the gauge potential. Once this relation is established, we show that it translates into the product of the decay constant $f$ and the instanton action $S$ derived from the same cycle but wrapped by a different brane. This allows us to conjecture that a fundamental axion in a consistent $4d$ EFT in a quasi-dS background should satisfy
\begin{equation}\label{eq:aFLbound}
    S f \gtrsim \sqrt{M_P H} \sim V^{1/4}\fstop
\end{equation}
Together with the axionic version of the WGC, it provides a window of allowed parameters, as we show in \cref{fig:multiFL} in the case of multiple axions. 

Let us note in passing that Eq.~\eqref{eq:aFLbound} is bounding from below the values of the decay constant. Given that $S>1$ for consistency, our bound is also stating that the limit $f\rightarrow 0$ for a fundamental axion is not allowed in the EFT. As pointed out in~\cite{Reece:2018zvv,Heidenreich:2021yda}, for a fundamental axion the point where the decay constant shrinks to zero corresponds to an infinite volume limit, where the effective description breaks down and hence cannot be reached in a consistent EFT. Hence, the aFL bound corroborates this statement, providing a reason why such limits should not occur.

By construction, the bound applies in situations of quasi-dS spacetime, thus providing interesting applications and consequences nowadays and during the inflationary epoch. For the latter case, we studied the consequences of aFL on axion monodromy and blow-up inflation. When both aFL and the WGC are taken into account, the parameter space for the amplitude and the frequency
of periodic instanton corrections get constrained. We find that the interplay between WGC and FL bounds constraints the range of allowed oscillatory contributions to the CMB power spectrum and resonant non-Gaussianities. Instead, for blow-up inflation the model constraints arise from the volumes ratio underlying WGC and FL conjectures. We show that, for standard parameter values needed for this model to work, the inflationary observables are out of reach of current experiments. Finally, we apply the aFL bound to axion-photon coupling and find that the mixing angle is bounded from below, ruling out a large portion of the parameter space currently used for model building. Furthermore, recently proposed values for the mixing angle derived from CMB constraints agree with our bound.

Given that the original FL bound~\cite{Montero:2019ekk,Montero:2021otb} was derived from charged black holes nearly as big as the dS horizon, we tried to study similar situations for the case of axion charges and their dual $H_3$ field strengths. However, we saw that it is not possible to give a direct derivation of aFL from the evaporation and subsequent discharge of axionically charged black holes~\cite{Bowick:1988xh}. The crucial issue is that the string networks produced as the axionic black hole discharges have equation of state $w=-1/3$, which does not result in spacetime crunching and becoming singular. On the contrary, they help the accelerated expansion.\footnote{More generally, one expects that whether a direct black hole argument for a form of FL in $D$ spacetime dimensions for objects charged under a $p$-form gauge fields works centrally depends on whether the equation of state of said objects drives expansion or contraction of the spacetime.} Similarly, we were unable to directly derive aFL by considering axionically charged euclidean wormholes.

It would be interesting to study whether experimental constraints on the current accelerated expansion could put bounds on the amount of strings emitted or on the time the black hole has to discharge, thus maybe allowing for a different and complementary argument for our aFL bound.  We leave these interesting questions for future work.

\section*{Acknowledgments}
We thank Jacob M. Leedom, Alessandro Mininno, Vincent Van Hemelryck and Thomas Van Riet for valuable discussion. V.G. and A.W. are supported by the ERC Consolidator Grant STRINGFLATION under the HORIZON 2020 grant agreement no. 647995. N.R. is supported by the Deutsche Forschungsgemeinschaft under Germany's Excellence Strategy - EXC 2121 ``Quantum Universe'' - 390833306.

\bibliographystyle{JHEP}
\bibliography{ref}

\end{document}